\documentclass[english,12pt]{article}
\usepackage{amsxtra,amssymb,amsfonts,amsthm,amsmath}
\usepackage{graphicx}
\usepackage[latin9]{inputenc}
\usepackage{mathrsfs}
\usepackage[letterpaper]{geometry}
\usepackage{setspace}
\usepackage{babel}
\usepackage{float}
\usepackage{cancel}

\makeatletter

\def\mathscr{\mathfrak}
\def\ds{\displaystyle}

\def\l{\lambda}

\def\d{\partial}

\def\const{\text{const}}
\def\sign{\text{sign}}
\def\ds{\displaystyle}
\def\dd{\partial}
\def\a{\alpha}
\def\b{\beta}
\def\g{\gamma}
\def\d{\delta}
\def\e{\varepsilon}
\def\ee{\epsilon}
\def\l{\lambda}
\def\o{\omega}
\def\t{\theta}

\date{}

\makeatother

\begin{document}

\title{Realization of Nonholonomic Constraints\\ and\\ Singular Perturbation Theory\\ for Plane Dumbbells}

\author{Sergiy Koshkin\\
Department of Mathematics and Statistics\\
University of Houston-Downtown\\
1 Main Street\\
Houston, TX 77002\\
e-mail: koshkins@uhd.edu \and 
Vojin Jovanovic\\
Systems, Implementation \& Integration\\
Smith Bits, A Schlumberger Co.\\
1310 Rankin Road\\
Houston, TX 77032\\
e-mail: fractal97@hotmail.com}

\maketitle

\newpage

\begin{abstract} We study the dynamics of pairs of connected masses in the plane, when nonholonomic (knife-edge) constraints are realized by forces of viscous friction, in particular its relation to constrained dynamics, and its approximation by the method of matching asymptotics of singular perturbation theory when the mass to friction ratio is taken as the small parameter. It turns out that long term behaviors of the frictional and constrained systems may differ dramatically no matter how small the perturbation is, and when this happens is not determined by any transparent feature of the equations of motion. The choice of effective time scales for matching asymptotics is also subtle and non-obvious, and secular terms appearing in them can not be dealt with by the classical methods. Our analysis is based on comparison to analytic solutions, and we present a reduction procedure for plane dumbbells that leads to them in some cases. 

\bigskip

\end{abstract}

\newpage

\section*{Introduction}\label{s0}

Nonholonomic mechanics is experiencing something of a revival in recent decades, see \cite{BM} for a historical perspective, and \cite{LM,Mar} for some of the new themes. In this paper we look at approximating nonholonomic constraints by anisotropic viscous friction for plane "dumbbells", pairs of masses joined by a weightless connecting mechanism of some sort \cite[2.1]{Green}. They are traditional toy models for exploring various mechanical effects and behaviors, see e.g. \cite{Ben,Green,Zek}, and a natural starting point for gaining insight into approximations of nonholonomic systems. 

The idea of de-idealizing nonholonomic constraints by "realizing" (replacing) them with forces of viscous friction goes at least as far back as Carath\'eodory \cite{Car}. Carath\'eodory considered the case of the Chaplygin sleigh and concluded that motions of the skidding sleigh, with knife-edge constraints replaced by viscous friction forces, did not converge to the constrained motions. After the development of perturbation theory for systems of ordinary differential equations in the 1950-60s Fufaev \cite{Fuf} analyzed the skidding sleigh  in terms of slow and fast motions, and showed that convergence does in fact take place for $t>0$; this is presented in his book with Neimark \cite[IV.3]{NF}. Both the Chaplygin and the 
Carath\'eodory-Fufaev sleighs are equivalent to dumbbells of the type we consider. 

In 1981 Brendelev  \cite{Bren} and Karapetian \cite{Kar}, see also \cite[1.6]{AKN}, \cite{Koz}, generalized Fufaev's analysis to general systems with linear velocity constraints. If one realizes them by viscous forces then in the limit of infinite friction frictional motions converge to constrained motions for positive time. The idea is that the viscous system quickly evolves towards the "slow manifold" of the constrained system (fast motion), and then remains in its vicinity (slow motion). More recently, Eldering \cite{Eld} clarified the nature of convergence to nonholonomic dynamics by interpreting the results of Brendelev and Karapetian in terms of the geometric theory of singular perturbations, which goes back to Fenichel \cite{Fen}.
Also recently, Deppler et al. considered  realization of constraints by more general viscoelastic forces, which may provide a closer approximation of the actual physics \cite{Dep}.

But analysis of concrete examples in light of what the general theory implies for them is hard to come by, in particular it turns out that the convergence involved does not guarantee the kind of approximating behavior one might expect. The general results provide convergence on a finite time interval, but the relations between fast and slow motions, estimation of their time scales, and long term behavior remain largely out of the picture. A quantitative account of transient effects occurring in such realizations was developed in the theory of singular perturbations by O'Malley and Vasil'eva, see e.g. \cite[Ch. 8]{Ver}, by using perturbative expansions and two time scales. In applications their approach came to be known as the method of matching asymptotics. Unfortunately, unlike applications to celestial mechanics and electrical engineering \cite{Holm, Ver}, biology \cite{Hek} and chemistry \cite{ZKK}, applications to nonholonomic dynamics appear to be understudied and underappreciated in the literature. We will show that the choice of appropriate time scales is non-trivial even in the simplest cases, complicating application of matching asymptotics, and the long term behaviors of nonholonomic systems and their realizations may differ. Moreover, we encounter interesting situations, where secular terms appear in perturbative expansions of equations that are non-oscillatory, so the standard methods for eliminating them, such as averaging  \cite[3.2]{Holm}, \cite[Ch.11]{Ver}, do not apply.

In this paper we first present a general approach for reducing the equations of motion of plane dumbbells to a form that simplifies analysis of their solutions (Section \ref{s1}). This involves representing positions of the masses as complex numbers, and manipulating equations in complex form before specific generalized coordinates are selected to take advantage of their symmetries and integrals of motion. We then focus on two examples, which we call the "cart sleigh" (Sections 2-3) and the "double spear" (Section 4), where the equations of motion turn out to be completely integrable, i.e. one can find analytic solutions. This allows us to compare approximate solutions obtained by the method of matching asymptotics to the exact ones, and assess their range of applicability precisely. This leads to a better understanding of the nature of convergence of motions in frictional realizations to the limiting motions of nonholonomic systems. 

The two examples we chose display two in some sense opposite behaviors in approximating nonholonomic systems in the limit of infinite friction. Whereas the skidding double spear "shadows" (approximates uniformly in time) its nonholonomic limit, the long term behavior of the skidding cart sleigh is completely different for arbitrarily large friction, despite the convergence of motions on finite time intervals. The latter effect seems to be underappreciated in the standard descriptions in terms of fast/slow motions, and limits the sense in which such descriptions can be "trusted". 

Our analysis is based on comparisons to analytic solutions. Discriminating between these two cases based on the equations of motion directly, and finding effective methods for approximating long term behavior of frictional solutions emerge as interesting open problems. Our conclusions are summarized in Section 5.

\section{Holonomic and nonholonmic dumbbells}\label{s1}

\begin{figure}[H]
\begin{centering}
\includegraphics[scale=1.2]{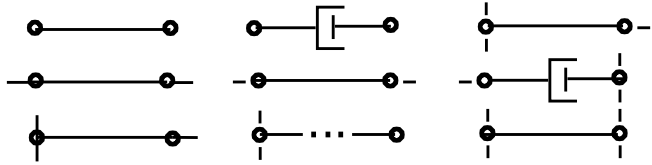}
\par\end{centering}
\caption{\label{Dumbbells} Dumbbell diagrams. Bullets represent masses, protruding short segments indicate directions of knife-edges, breaks with dots indicate telescoping rods, split off short segments indicate approximate realization of knife-edge constraints by viscous friction. The depiction of viscously damped rods by a `dashpot' is standard.}
\end{figure}

By a dumbbell we understand a pair of masses joined by a weightless connecting mechanism. The simplest examples are rigid rods, telescoping rods with or without damping, or springs with zero or non-zero equilibrium lengths. The connection is meant to physically implement forces between the masses acting along the segment connecting them. The masses themselves can be thought of as small balls free to move without friction along the plane, or as small balls mounted on knife-edges that constrain directions of their instantaneous velocities. Since we are interested in realization of such constraints strict constraints may be replaced with forces, e.g. of viscous friction, that approximate them. The knife-edges produce non-holonomic constraints linear in velocities, and we will only consider cases where they are either perpendicular or parallel to the connecting segments, although our approach generalizes to cases where they are attached at some other fixed angle. We only consider dumbbell motion in the plane. Schematic diagrams for various types of dumbbells are presented on Fig.\ref{Dumbbells}.
We start by describing how the equations of motion are derived in vector form. Let $r_1,r_2$ be the vectors of positions, and $m_1$, $m_2$ be the masses of a dumbbell, then the kinetic energy is $T=\frac12m_1|\dot{r}_1|^2+\frac12m_2|\dot{r}_2|^2$. For simplicity we assume that the masses are always equal, $m_1=m_2=m$, so $T=\frac{m}2(|\dot{r}_1|^2+|\dot{r}_2|^2)$. Perpendicular knife-edge is represented by the constraint $\dot{r}_i\perp(r_1-r_2)$, or in terms of the dot product $\dot{r}_i\cdot(r_1-r_2)=0$. For the parallel knife edge $\dot{r}_i\parallel(r_1-r_2)$. To give a dot product form it is convenient to introduce a linear transformation $J=\begin{pmatrix}0&-1\\1&0\end{pmatrix}$ that rotates every vector by $90^\circ$ counterclockwise, note that $J^*=J^{-1}=-J$. The constraint can then be represented by $\dot{r}_i\cdot J(r_1-r_2)=0$. The rigid rod constraint is holonomic, $|r_1-r_2|=\const$, but it is often convenient to work with its differential consequence obtained by differentiating $(r_1-r_2)\cdot(r_1-r_2)=\const$, namely $(\dot{r_1}-\dot{r_2})\cdot(r_1-r_2)=0$. 

The forces of viscous friction realizing knife-edges are described by the Rayleigh dissipation function \cite[2.4]{Green}. For the knife-edge attached to the $i$-th mass perpendicular to the connecting rod the Rayleigh function is $R_i:=\frac{c}2|\text{pr}_{r_1-r_2}\dot{r}_i|^2$ , where $c>0$ is the friction coefficient, and $\text{pr}_ab$ denotes the orthogonal projection of $b$ to $a$. Since 
$\text{pr}_ab=\frac{b\cdot a}{|a|^2}a$ we have explicitly 
$$
R_i:=\frac{c}2\frac{|\dot{r}_i\cdot(r_1-r_2)|^2}{|r_1-r_2|^2}\,.
$$
For parallel knife edges $r_1-r_2$ is replaced by $J(r_1-r_2)$, note that $|J(r_1-r_2)|=|r_1-r_2|$. Similarly, for damped rods the Rayleigh function is 
$$
R=\frac{c}2\,|\text{pr}_{r_1-r_2}(\dot{r_1}-\dot{r_2})|^2=\frac{c}2\,\frac{|(\dot{r_1}-\dot{r_2})\cdot(r_1-r_2)|^2}{|r_1-r_2|^2}\,.
$$
In the absence of constraints the equations of motion in vector form are given by 
\begin{equation}\label{LagRayl}
\frac{d}{dt}\frac{\dd L}{\dd\dot{r}_i}-\frac{\dd L}{\dd r_i}+\frac{\dd R}{\dd\dot{r}_i}=0\,.
\end{equation}
where $\frac{\dd}{\dd a}$ stands for the partial gradient along the vector $a$, and $L:=T-U$ is the Lagrangian function. A straightforward computation shows that $\frac{\dd|Aa|^2}{\dd a}=2A^*Aa$, where $A$ is a linear transformation, and this along with the chain rule is enough to differentiate all the functions that we need in this paper. In particular, 
\begin{align}\label{Deriv}
&\frac{\dd}{\dd\dot{r}_i}\,\frac{m}2(|\dot{r}_1|^2+|\dot{r}_2|^2)=m\dot{r}_i\\
&\frac{\dd}{\dd\dot{r}_i}\,\frac{c}2|\text{pr}_{r_1-r_2}(\dot{r}_i)|^2=c\,\text{pr}_{r_1-r_2}(\dot{r}_i)
=c\,\frac{\dot{r}_i\cdot(r_1-r_2)}{|r_1-r_2|^2}(r_1-r_2),\notag
\end{align}
because $\text{pr}_a^*=\text{pr}_a^{2}=\text{pr}_a$ for orthogonal projections.

The constraints are handled according to d'Alembert's principle of zero virtual work. For the constraints linear in velocities that we are considering the following rule (sometimes called Jourdain's principle \cite{Bah}, \cite[2.4]{Green}) suffices: take the variation of the constraint with respect to velocities, multiply it by a Lagrange multiplier, and add the term in the dot product with $\d\dot{r_i}$ to the $i$-th equation in \eqref{LagRayl}. For example, to implement the differentiated rigid rod constraint we take the variation:
$$
\d\big((\dot{r_1}-\dot{r_2})\cdot(r_1-r_2)\big)=(r_1-r_2)\cdot(\d\dot{r_1}-\d\dot{r_2})\\
=(r_1-r_2)\cdot\d\dot{r_1}-(r_1-r_2)\cdot\d\dot{r_2}\,.
$$
Thus, we will add $\l(r_1-r_2)$ to the equation for $r_1$, and $-\l(r_1-r_2)$ to the one for $r_2$. This works even for non-linear constraints \cite{Bah}. 

The reasons we prefer to write equations in vector form initially rather than go straight to the generalized coordinates, as e.g. in \cite{Ben}, are twofold. Vector equations provide geometric insight and manifest symmetries that are often lost once some specific choice of generalized coordinates is made. Second, it is not always immediately clear what generalized coordinates are most beneficial for analyzing or solving the system. We shall see that when working from vector equations such choices often "suggest themselves". This can be seen as an intuitive version of non-holonomic reduction without the abstract formalism and technicalities of the general case \cite{Koi}. An important ingredient in this reduction process is the observation that vectors in the plane can be interpreted as complex numbers. Then applying $J=\begin{pmatrix}0&-1\\1&0\end{pmatrix}$ to a vector corresponds to multiplying it by the imaginary unit $i$, and the vector equations turn into scalar equations for complex numbers. The dot product also has a simple expression in terms of complex multiplication: $z\cdot w=\text{Re}[z\overline{w}]=\text{Re}[\overline{z}w]$. Complex numbers can be represented in Cartesian, exponential and polar forms, which provides a rich selection of real variables to choose from for generalized coordinates, and since complex numbers can be also multiplied and divided this selection is enriched by applying the idea to ratios of the original vectors and their combinations. 

\section{Cart sleigh}\label{s2}

Let us illustrate the outline of the previous section with a simple example of a telescopic rod dumbbell with perpendicular knife-edges. One can think of this dumbbell as a toy model of a sleigh with short sharp rails placed on ice (or a pair of skates affixed to a rod). If the knife-edges are replaced with wheels one gets a common simple model of a two-wheeled cart \cite[III.3]{NF}; we therefore call this dumbbell the cart sleigh. 
\begin{figure}[H]
\begin{centering}
\includegraphics[scale=1.2]{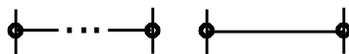}
\par\end{centering}
\caption{\label{TeleCartSleigh} Cart sleigh with telescopic and rigid rods.}
\end{figure}
The Lagrangian of the cart sleigh is just the kinetic energy $L=\frac{m}2(|\dot{r}_1|^2+|\dot{r}_1|^2)$, the knife-edges are implemented by constraints $\dot{r}_1\cdot(r_1-r_2)=0$ and $\dot{r}_2\cdot(r_1-r_2)=0$. Even before writing the equations of motion subtracting the second constraint from the first gives us $(\dot{r_1}-\dot{r_2})\cdot(r_1-r_2)=0$, which is none other than the differential form of the rigid rod constraint. In other words, in this case a telescopic rod behaves as if it was rigid! Now $\frac{d}{dt}\frac{\dd L}{\dd\dot{r}_i}=m\ddot{r}_i$, and applying Jourdain's principle to each constraint we get the equations of motion
$$
\begin{cases}m\ddot{r}_1+\l_1(r_1-r_2)=0\\ m\ddot{r}_2+\l_2(r_1-r_2)=0\,.\end{cases}
$$
Since we have $r_1-r_2$ in both equations it seems natural to introduce a new vector variable $r_{12}:=\frac12(r_1-r_2)$ (the reason for $\frac12$ will become clear shortly). Subtracting we get a self-contained equation $m\ddot{r}_{12}+(\l_1-\l_2)r_{12}=0$ with the transformed constraint $\dot{r}_{12}\cdot r_{12}=0$. We now need a complementary variable, for which $r:=\frac12(r_1+r_2)$ is a natural candidate. The $\frac12$ coefficients ensure that both $r_1,r_2$ can then be recovered without fractions: 
$r_1=r+r_{12}$, $r_2=r-r_{12}$, and $r$ has the physical interpretation of being the center of mass of the dumbbell (and the midpoint of the rod). Adding the equations and the constraints we get $m\ddot{r}+(\l_1+\l_2)r_{12}=0$ and $\dot{r}\cdot r_{12}=0$, which directly tells us that the velocity of the center is always perpendicular to the rod. Since the Lagrange multipliers are also unknown we might as well set $\l_{12}:=\l_1-\l_2$, $\l:=\l_1+\l_2$, which results in the system:
\begin{equation}\label{CartVecEq}
\begin{cases}m\ddot{r}_{12}+\l_{12}r_{12}=0\\
m\ddot{r}+\l r_{12}=0\\
\dot{r}_{12}\cdot r_{12}=0,\,\,\dot{r}\cdot r_{12}=0\,.\end{cases}
\end{equation}
Some information about the motion can be extracted directly from this vector system. Taking dot product of the first equation with $\dot{r}_{12}$, and the second one with $\dot{r}$ we get right away that $\ddot{r}_{12}\cdot\dot{r}_{12}=\frac{d}{dt}\frac12|\dot{r}_{12}|^2=0$, so $|\dot{r}_{12}|=\const$, and similarly $|\dot{r}|=\const$, i.e. both the center, and each mass relative to the center (which is what $\pm r_{12}$ represent) move with constant speeds.

Now it is time for complex notation. Since $|r_{12}|=\rho=\const$ it is natural to choose exponential form for 
$r_{12}:=\rho e^{i\t}$, then 
$$
\dot{r}_{12}=i\dot{\t}\rho e^{i\t},\text{ and }
\ddot{r}_{12}=(-\dot{\t}^2+i\ddot{\t})\rho e^{i\t}\,.
$$
With this notation the first equation in \eqref{CartVecEq} is
$$
m(-\dot{\t}^2+i\ddot{\t})\rho e^{i\t}+\l_{12}\rho e^{i\t}=0\,.
$$
As $\l_{12}$ is real-valued we get from separating real and imaginary parts
$$
\dot{r}_{12}=i\dot{\t}\rho e^{i\t},\hspace{5em}
\ddot{r}_{12}=(-\dot{\t}^2+i\ddot{\t})\rho e^{i\t}\,,
$$
Since $\t$ enters only through $\dot{\t}$ it is natural to set $\o:=\dot{\t}$, so that $\l_{12}=m\o^2$ and $\dot{\o}=0$, i.e. $\o=\const$. The dumbbell rotates with constant angular velocity (this is expected from the constant speed and length of $r_{12}$).

\begin{figure}[H]
\vspace{-0.1 in}
\begin{centering}
\includegraphics[scale=0.3]{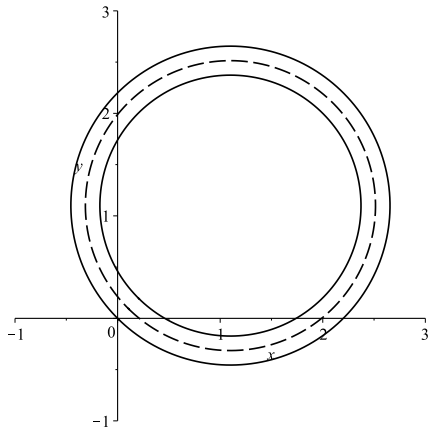}
\par\end{centering}
\caption{\label{p1concentric} Cart sleigh moving in a circle. Solid lines are the trajectories of the knife-edges, dashed line is the trajectory of the center of mass.}
\end{figure}

We also notice that $r$ enters the equations only through $\dot{r}$, $|\dot{r}|=\const$, and $\dot{r}\perp r_{12}$. This means that $\dot{r}$ is a constant (real!) multiple of $ir_{12}$, $\dot{r}=i\b r_{12}$ $\ddot{r}=i\b\dot{r}_{12}=-\b\o\rho e^{i\t}$
reducing the second equation to $-m\b\o\rho e^{i\t}+\l\rho e^{i\t}=0$, so $\l=\b m\o$. The values of $\b$ and $\o$ can be found from initial conditions. We now have ${r}_{12}=\rho e^{i(\t_0+\o t)}$, and $\dot{r}=i\b\rho e^{i(\t_0+\o t)}$, which yields
\begin{multline}\label{CartPos}
r(t)=r(0)+\int_0^ti\b\rho e^{i(\t_0+\o t)}dt=r(0)+\frac{\b\rho}{\o}e^{i(\t_0+\o t)}\Big|_0^t\\
=r(0)+\b\rho e^{i\t_0}\,\frac{e^{i\o t}-1}{\o}
=r(0)-\frac{\b\rho}{\o}e^{i\t_0}+\frac{\b\rho}{\o}e^{i(\t_0+\o t)}
\end{multline}
This means that the center of mass uniformly rotates along a circle with the center $r(0)-\b\rho e^{i\t_0}/\o=r(0)-\dot{r}(0)/\o$ of radius $\dot{r}(0)/\o$, while the masses uniformly rotate along a circle centered at the moving point of the radius. Their resulting motion is therefore epicyclic (as in geocentric astronomical models), with $r$ tracing the deferent, and $r_{12}$ the epicycle. Due to the equal frequencies and phases this is a very special case of epicyclic motion however. For the masses the resulting motion is 
$$
r_{1,2}=r\pm r_{12}=r(0)-\dot{r}(0)/\o
+(\b/\o\pm1)\rho e^{i(\t_0+\o t)},
$$ 
i.e. the resulting trajectory is itself a circle with the same center as the deferent, and the radius smaller/larger by the radius of the epicycle. Thus, the center of mass and both masses move along concentric circles with constant angular velocities, Fig. \ref{p1concentric}. The above discussion applies to the case $\o\neq0$, the $\o=0$ case can be obtained by taking the limit $\o\to0$ in \eqref{CartPos}, which describes uniform rectilinear motion in the direction of the initial push. 

Of course, for the cart sleigh one can predict the final answer using physical intuition, but things usually do not work out so simply and neatly. Still, in a number of cases the reduction procedure outlined above leads to an analytic solution of dumbbell equations, and even when it does not at least some qualitative features and integrals of motion can be found as a matter of course. Moreover, when constraints are realized by large viscous forces, which is our point of interest, equations are reduced to a form where asymptotic methods of perturbation theory can be applied fruitfully by encapsulating friction coefficients into small parameters, with nonholonomic limits representing the unperturbed motion. The viscous motions then reveal some surprises already in the case of the cart sleigh, as we show next.

\section{The sleigh skids}\label{s3} 

Consider what happens if we kick the cart sleigh in a way that produces initial velocities inconsistent with the constraints? The formal answer is that we can not, literally. The equations of motion for a constrained system with initial conditions not conforming to the constraints have no solutions. At least no classical solutions, but it is also unclear what 
"non-classical solutions" might mean here. This is not to say that the sleigh simply can not start moving, that would still be a solution, the trivial one, while no solution means that the theory literally predicts nothing in this case. But under the usual intuition of mechanical idealizations a "kick" amounts to an "infinite" impulse force imparting finite instantaneous velocity, and it can act in any direction. So ideally speaking we "can" in fact kick the cart sleigh in a non-conforming way, and it "should" do something. One could suggest that the "infinite" reaction force in response to the "kick" cancels its component normal to the constraints, so it instantly turns initial conditions into conforming ones. But does this intuition reflect the behavior under what the "kick" is supposed to approximate?

In this section we will investigate what happens if the constraint idealization is relaxed in the spirit of Carath\'eodory and Fufaev, i.e. if skidding is allowed. We realize the constraints with large viscous friction, and discuss if it makes sense to talk about "solutions" to non-holonomic equations of motion (as limits) when the initial values do not conform to the constraints. After deriving the equations of motion we first apply the popular method of matching asymptotics to it, which is a standard method in singular perturbation theory \cite[3.2]{Holm}, \cite[Ch.8]{Ver}, but is rarely if ever used 
in the context of non-holonomic dynamics. Part of our goal is to call attention to it, and to highlight the advantages and the challenges of its application. Then we derive the analytic solution and compare it to the matched asymptotic expansion, which illustrates both its strengths and limitations.
\begin{figure}[H]
\begin{centering}
\includegraphics[scale=1.2]{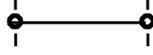}
\par\end{centering}
\caption{\label{CartSleigh} Skidding cart sleigh. Split off short segments indicate approximate realization of knife-edge constraints by viscous friction.}
\end{figure}

In the case of the cart sleigh it made no difference whether we allowed the connecting rod to telescope or made it rigid, but when the knife-edges are allowed to skid it does make a difference. We pick the rigid rod realization, which is simpler. The new elements are the viscous forces, which replace the knife-edge constraints and are given by the Rayleigh function 
$$
F=\frac{c}2\left(\frac{|\dot{r_1}\cdot(r_1-r_2)|^2}{|r_1-r_2|^2}
+\frac{|\dot{r_2}\cdot(r_1-r_2)|^2}{|r_1-r_2|^2}\right), 
$$
and the explicit rigid rod constraint in the differential form
$(\dot{r_1}-\dot{r_2})\cdot(r_1-r_2)=0$. The vector equations of motion are obtained as before (see \eqref{Deriv} for taking the derivative of the Rayleigh function): 
$$
\begin{cases}m\ddot{r}_1+\l(r_1-r_2)+c\,\ds{\frac{\dot{r}_1\cdot(r_1-r_2)}{|r_1-r_2|^2}}(r_1-r_2)=0\\ 
m\ddot{r}_2-\l(r_1-r_2)+c\,\ds{\frac{\dot{r}_2\cdot(r_1-r_2)}{|r_1-r_2|^2}}(r_1-r_2)=0\,.\end{cases}
$$
After the familiar substitution $r_{12}:=\frac12(r_1-r_2)$, $r:=\frac12(r_1+r_2)$, 
adding and subtracting the equations while taking into account that $\dot{r}_{12}\cdot r_{12}=0$ results in
$$
\begin{cases}m\ddot{r}_{12}+2\l r_{12}=0\\
m\ddot{r}+c\,\ds{\frac{\dot{r}\cdot r_{12}}{|r_{12}|^2}\,r_{12}}=0\,.\end{cases}
$$
Interpreting $r,r_{12}$ as complex numbers we set $r_{12}=\rho e^{i\t}$ with $\rho=\const$, and the first equation tells us that $\o:=\dot{\t}=\const$ and $\l=m\o^2/2$ as in the constrained case. The first equation can now be solved explicitly, and gives $r_{12}(t)=r_{12}(0)e^{i\t}=\rho e^{i(\t_0+\o t)}$. Unlike before, however, we do not have that $\dot{r}\perp r_{12}$  or that $|\dot{r}|=\const$. Nonetheless, since we expect $\dot{r}=i\b r_{12}$ in the limit let us set $\dot{r}=z r_{12}$, where $z:=\a+i\b$ is now a variable, and complex valued. Then 
$$
\dot{r}\cdot r_{12}=\text{Re}[(\a+i\b)\rho e^{i\t}\,\overline{\rho e^{i\t}}]=\a\rho^2,\ \ 
\ddot{r}=[(\dot{\a}-\o\b)+i(\dot{\b}+\o\a)]\rho e^{i\t}\,,
$$
so after separating the real and imaginary parts the second equation reduces to a pair of equations for $\a,\b$:
$$
\begin{cases}
\dot{\b}+\o\a=0\\
m(\dot{\a}-\o\b)+c\a=0\,.
\end{cases}
$$
Since we plan to investigate what happens when $c\to\infty$ it makes sense to introduce a small 
parameter $\e:=m/c$, and rewrite the system in the Tikhonov form \cite[Ch.8]{Ver}
\begin{equation}\label{CartSlow}
\begin{cases}\dot{\b}=-\o\a\\
\e\dot{\a}=-\a+\e\o\b\,.
\end{cases}
\end{equation}

This is the reduced system we will analyze. The system is linear and can be solved exactly, but as our goal is general insight let us disregard that for the moment, and apply a method that would work for non-linear systems as well. Setting $\e=0$ (i.e. $c=\infty$) gives $\a=0$ and $\dot{\b}=0$, i.e. $\b=\const$: this is the cart sleigh considered in the previous section, as expected. To get the next order of approximation in $\e$ we proceed as follows. Set $\a=\a^0+\e\a^1+\dots$, $\b=\b^0+\e\b^1+\dots$, so upon substitution into \eqref{CartSlow} we have
\begin{equation}\label{CartSlowEq}
\begin{cases}
\dot{\b}^0+\e\dot{\b}^1+\dots=-\o(\a^0+\e\a^1+\dots)\\
\e\dot{\a}^0+\e^2\dot{\a}^1+\dots=-\a^0-\e\a^1-\dots+\o(\e\b^0+\e^2\b^1+\dots)\,.
\end{cases}
\end{equation}
We then equate the terms in each order of $\e$.

The zero order terms simply reproduce what we found above: $\a^0=0$, $\dot{\b}^0=0$, and in the first order we get 
$$
\begin{cases}
\dot{\b}^1=-\o\a^1\\
\dot{\a}^0=-\a^1+\o\b^0\,.
\end{cases}
$$
Let $\b^0=B_0=\const$, then $\a^1=\o B_0$ and $\dot{\b}^1=-\o^2 B_0$, so $\dot{\b}^1=-\o^2 B_0t+B_1$, where $B_1$ is another constant. Thus, to the first order in $\e$ we find 
\begin{equation}\label{CartSlowSol}
\begin{cases}
\a=\e\o B_0+O(\e^2)\\
\b=B_0+\e(-\o^2 B_0t+B_1)+O(\e^2)\,.
\end{cases}
\end{equation}
We could try to find $B_0,B_1$ by using initial values for $\a,\b$, but that would be premature. The theory implies that this  
{\it regular expansion} is only valid for "slow motions", into which the system settles after a short transient period, so the integration constants are not directly related to the initial values.

To find the transient "fast motions" the standard approach is to introduce stretched time $\tau:=t/\e$, then 
$\frac{d}{dt}=\frac1{\e}\frac{d}{d\tau}$. Denoting derivatives with respect to $\tau$ by $'$ we transform \eqref{CartSlow} into 
\begin{equation}\label{CartFast}
\begin{cases}\b'=-\e\o\a\\
\a'=-\a+\e\o\b\,.
\end{cases}
\end{equation}
Expanding as in \eqref{CartSlowEq} produces to the first order in $\e$: 
\begin{equation}\label{CartFastEq}
\begin{cases}\b^{0\,\prime}=0\\
\a^{0\,\prime}=-\a^0
\end{cases}
\begin{cases}\b^{1\,\prime}=-\o\a^0\\
\a^{1\,\prime}=-\a^1+\o\b^0\,.
\end{cases}
\end{equation}
Thus, $\b^0=\const$, $\a^0=A_0e^{-\tau}$, and since these are fast motions we are now justified to find the integration constants from the initial values, i.e. 
\begin{equation}\label{CartFastConst}
\begin{cases}\b^0=\b(0)=:\b_0\\
\a^0=\a(0)e^{-\tau}=:\a_0\,e^{-\tau}\,.
\end{cases}
\end{equation}
We now turn to matching. The idea is that right after the initial kick the system undergoes the fast motion, and its asymptotic values of $\a,\b$ become the initial values for the subsequent slow motion. This finally allows us to connect the constants $B_0,B_1$ to the initial values.

Note that for the slow motions $\a^0=0$, which matches with the limit of the fast motion at temporal infinity, $\lim_{\tau\to\infty}\a_0\,e^{-\tau}=0$. The same type of matching for $\b^0$, which remains constant in the zero order during the fast motion gives $B_0=\b_0=\b(0)$. Since the initial values are accounted for in the zero order, in the first order we should set $\a^1(0)=\b^1(0)=0$. Solving \eqref{CartFastEq} we then have
\begin{equation}\label{CartFastSol}
\begin{cases}
\a^1=\b_0\o(1-e^{-\tau})\\
\b^1=-\a_0\o(1-e^{-\tau})\,.
\end{cases}
\end{equation}
When $\tau\to\infty$ this should match first order terms in \eqref{CartSlowSol} with $t\to0$. For $\a^1$ the match is automatic, and for $\b^1$ we determine $B_1=-\a_0\o$. Combining \eqref{CartSlowSol}, \eqref{CartFastConst} and  \eqref{CartFastSol} we get what is called the two timing or matched perturbative expansion to the first order in $\e$:
\begin{equation}\label{CartTwoSol}
\begin{cases}
\a=\a_0e^{-t/\e}+\e\o\b_0(1-e^{-t/\e})+O(\e^2)\\
\b=\b_0-\e\o\big(\b_0\o t+\a_0\o(1-e^{-t/\e})\big)+O(\e^2)\,.
\end{cases}
\end{equation}
The approach we outlined is the method of matching asymptotics mathematically justified by O'Malley and Vasil'eva in the case of singular perturbations. The expansion is valid as asymptotic expansion on a finite time interval $[0,T]$ with $T\sim O(1)$, 
see \cite[8.3]{Ver}. The last restriction on the length of time interval is not a formality as we will see. 

We can now attempt to use fast motion to define "generalized solutions" for constrained systems with non-conforming initial values. They should be the classical solutions with initial values replaced by the values "at the end" of the fast motion, that is by their limits as $\tau\to\infty$. In our example this amounts to replacing $(\a_0,\b_0)$ with $(0,\b_0)$. This is equivalent to the intuition of reaction forces instantly canceling the component of the initial "kick" orthogonal to the constraints. But it turns out that if we do so we can not count on the frictional solutions staying close to the constrained solutions beyond $T\sim O(1)$.
\begin{figure}[H]
\vspace{-0.2in}
\begin{centering}
\includegraphics[scale=0.5]{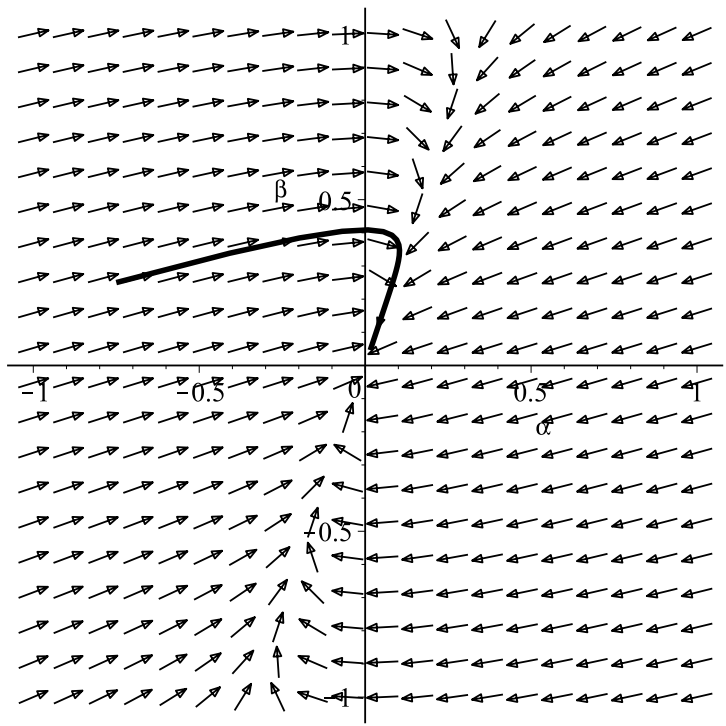}
\par\end{centering}
\caption{\label{1phase}. Phase portrait of \eqref{CartSlow} for initial conditions 
$r_1(0)=0$, $r_2(0)=2+2i$, $\dot{r}(0)=1+i/2$, $\o=3/2$ and $\e=1/5$.}
\end{figure}
To see this let us consider system \eqref{CartSlow} from a different perspective. The phase portrait of the system in the
$\a$-$\b$ plane along with a sample trajectory is shown on Fig.\ref{1phase}. As we already remarked, \eqref{CartSlow} is linear and can be solved exactly. Looking for solutions of the form $\b=e^{\xi t}$ we find $\a=-\frac{\xi}\o\,e^{\xi t}$, and the characteristic equation for $\xi$, namely $\e\xi^2+\xi+\o^2=0$. Its two roots are $\xi_{1,2}=-\frac{1\pm\sqrt{1-4\e^2\o^2}}{2\e}$, both real and negative for $\e\o<1/2$. The general solution is of the form $\b=b_1e^{\xi_1 t}+b_2e^{\xi_2 t}$, and enforcing initial values one finds:
\begin{multline*}
\a=(1-4\e^2\o^2)^{-1/2}\left(-\left(\frac{1-\sqrt{1-4\e^2\o^2}}{2}\a_0+\e\o\b_0\right)e^{-\frac{1-\sqrt{1-4\e^2\o^2}}{2\e}t}\right.\\
\left.+\left(\frac{1-\sqrt{1+4\e^2\o^2}}{2}\a_0-\e\o\b_0\right)e^{-\frac{1+\sqrt{1-4\e^2\o^2}}{2\e}t}\right)\!;
\end{multline*}
\vspace{-1em}
\begin{multline}\label{CartAnSol}
\b=(1-4\e^2\o^2)^{-1/2}\left(\left(\frac{1+\sqrt{1-4\e^2\o^2}}{2}\b_0-\e\o\a_0\right)e^{-\frac{1-\sqrt{1-4\e^2\o^2}}{2\e}t}\right.\\
\left.+\left(-\frac{1-\sqrt{1-4\e^2\o^2}}{2}\b_0+\e\o\a_0\right)e^{-\frac{1+\sqrt{1-4\e^2\o^2}}{2\e}t}\right)\!.
\end{multline}
The formulas are cumbersome, but one can see by inspection that for any $\e>0$ the exponents are strictly negative, so $\a(t),\b(t)\xrightarrow[t\to\infty]{}0$. Intuitively, this is apparent from the phase portrait, where the origin is a globally attractive equilibrium. Note that the constrained cart sleigh is represented by the $\b$-axis, with constrained dynamics consisting of fixed points. Recall that $\dot{r}=(\a+i\b)r_{12}$, and $r_{12}(t)=\rho e^{i\t}$, which means that $\dot{r}(t)\xrightarrow[t\to\infty]{}0$. Unlike for the constrained sleigh of the previous section the skidding sleigh's center of mass tends to rest asymptotically no matter how large the viscous friction is, and despite the fact that the friction forces quickly become "almost" orthogonal to $\dot{r}$, so should do "almost" no work against the overall motion of the sleigh.  Thus, the "generalized solutions" to the constrained equations will drift further and further apart from the solutions for $\e>0$ as time goes on despite the latter getting closer and closer to the slow manifold of the constrained solutions $\a=0$. This is because they are approaching it by approaching the origin, while for the constrained motion $\b$ remains constant, and so does not approach $0$.
\begin{figure}[H]
\vspace{-0.22in}
\begin{centering}
\includegraphics[scale=0.3]{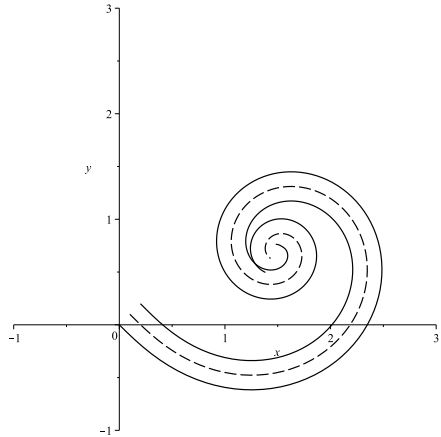}
\par\end{centering}
\caption{\label{p1spiral} Skidding cart sleigh moving. Solid lines are the trajectories of the knife-edges, dashed line is the trajectory of the center of mass.}
\end{figure}
As one can see from Fig. \ref{p1spiral} for a typical trajectory masses move along winding in spirals, and the center of mass asymptotically approaches a point, around which the masses spin with angular velocity $\o$. This should be compared to Fig. \ref{p1concentric} depicting the motion of a constrained sleigh, where the masses and the center of mass are moving along concentric circles. Note that even the skidding sleigh does not come to a stop despite losing energy to the constraints. Instead, the system "finds" an asymptotic motion where constraints no longer do any work, and the residual energy is conserved. In this case it can be attributed to the idealization of the "infinitesimal length" knife-edges. Generally speaking, the energy is conserved in constrained motions, so as long as there exist non-trivial ones we can expect non-trivial asymptotic motions that conserve energy.

An issue with matching asymptotics is already apparent from the second equation in \eqref{CartTwoSol}, which includes a term proportional to $t$ in the first order of $\e$. If we take this term at face value then for large $t$ our $\b$, and hence $\dot{r}$, would become arbitrarily large as $t\to\infty$. Terms of this nature first appeared in celestial mechanics when approximating the motion of the planets, and came to be called "secular terms". Their presence indicates that \eqref{CartTwoSol} can not be relied upon for large $t$. At the end of 19th century Lindstedt and Poincare developed a method for "eliminating" secular terms, i.e. constructing expansions that do not contain them, and therefore have a chance of holding uniformly in time. Unfortunately, their approach, and the method of averaging developed later, rely on detecting resonances and avoiding them in expansions, which is specific to equations that in their unperturbed form describe something close to periodic motions \cite[3.1]{Holm}, \cite[10.1]{Ver}. But \eqref{CartSlow} is non-oscillatory for $\e=0$ or small $\e>0$, so the standard apparatus of averaging does not apply. In fact, this is a general feature one can expect from perturbative expansions for realization of non-holonomic constraints by forces of viscous friction: all the pain of secular terms, no benefit of averaging.

We can sidestep the difficulty in this case only because we have the exact analytic solution. To better understand long term behavior let us expand \eqref{CartAnSol} to the first order in $\e$ using that 
$\sqrt{1-4\e^2\o^2}=1-2\e^2\o^2+O(\e^4)$:
\begin{flalign}\label{CartAnExp}
\a&=\a_0e^{-t/\e+\o^2\e t}+\e\o\b_0e^{-\o^2\e t}+O(\e^2);\notag\\
\b&=(\b_0-\e\o\a_0)e^{-\o^2\e t}+\e\o\a_0e^{-t/\e+\o^2\e t}+O(\e^2)\\
&\hspace{10em}=\b_0e^{-\o^2\e t}+\e\o\a_0(e^{-t/\e+\o^2\e t}-e^{-\o^2\e t})+O(\e^2)\,.\notag
\end{flalign}
This expansion, unlike \eqref{CartTwoSol}, is not just $O(1)$, one can show that it converges to the exact solution for all 
$t>0$ when $\e\to0$. The time scales manifestly present in \eqref{CartAnExp} are the stretched time $t/\e$ of fast motion, and the compressed time $\e t$ of (very) slow motion. In the sense of matching asymptotics, which relies on the apparent form of the terms, the ordinary slow time scale $t$ is not present at all! When we "forced" it into the expansion \eqref{CartTwoSol} by naively using matching asymptotics the result was a secular term that restricted its validity. This does not preclude \eqref{CartTwoSol} from holding for $t\sim O(1)$ of course, but its form is quite misleading the long term. The value of $\b$ drifts to $0$ on the very slow time scale of $\e t$, so on $t/\e$ and even $t$ scales it registers as staying constant at zero order. The secular term emerges in the first order to compensate for the drift. 
\begin{figure}[H]
\vspace{-0.1in}
\begin{centering}
(a) \includegraphics[scale=0.3]{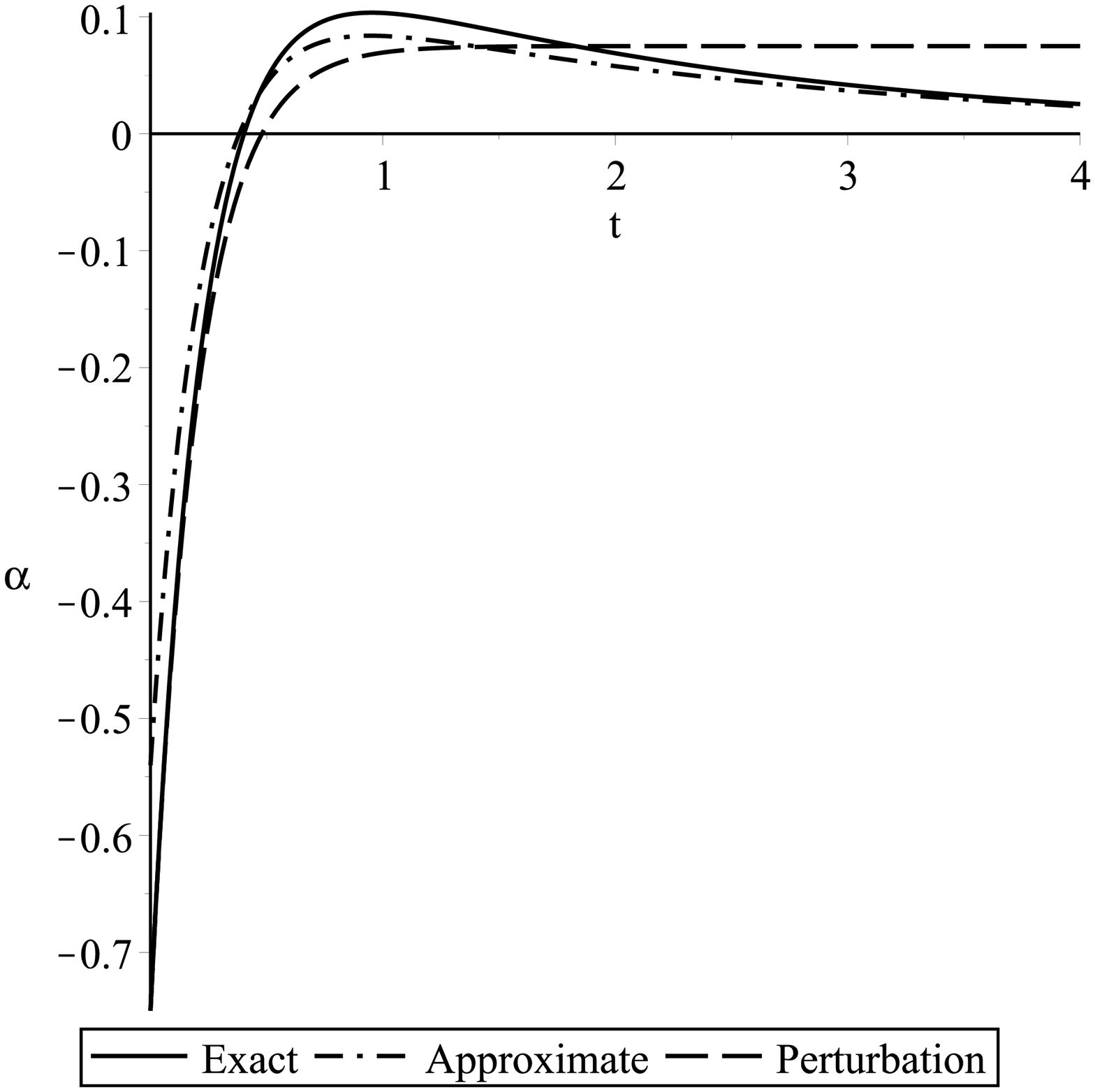}
(b) \includegraphics[scale=0.3]{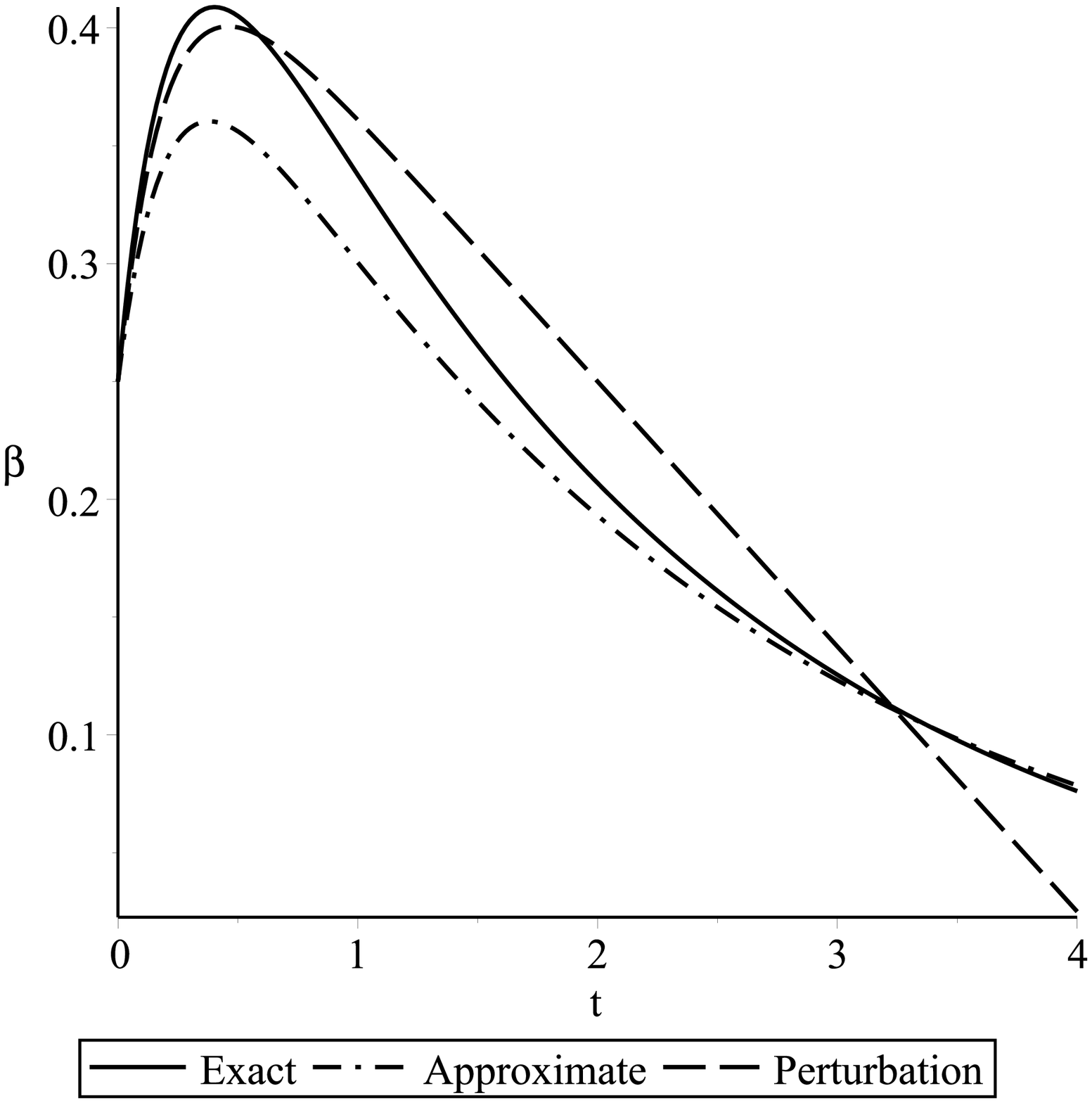} 
\par\end{centering}
\caption{\label{1alpha} Graphs of (a) $\a$ and (b) $\b$ as functions of time for initial conditions 
$r_1(0)=0$, $r_2(0)=2+2i$, $\dot{r}(0)=1+i/2$, $\o=3/2$ and $\e=1/5$.}
\end{figure}
Figure \ref{1alpha} displays graphs of the exact solution \eqref{CartAnSol}, its approximation \eqref{CartAnExp}, and the two timing perturbative expansion \eqref{CartTwoSol}, both to the first order. As expected, the latter remains close to the analytic solution for small 
$t$, but then deviates from it significantly.

Note also that even the analytic solution \eqref{CartAnSol} is only valid if $\e\o<1/2$. When $\o>1/2\e$, i.e. if the sleigh is made to spin fast enough initially, it is seen from the formula that the behavior changes qualitatively. The square root $\sqrt{1-4\e^2\o^2}$ becomes imaginary, and $\b$ undergoes damped oscillations with frequency $\sqrt{\o^2-1/4\e^2}$. Their amplitudes are multiplied by $e^{-t/2\e}$ however, so they all but disappear after the fast motion. For such initial values both \eqref{CartTwoSol} and \eqref{CartAnExp} are misleading even in the zero order, as they "predict" constant $\b^0$ during the fast motion, rather than damped oscillations. This is indicated by the presence of $\e\o$ terms in the first order, which then will not be "small". This illustrates the issue discussed by Holmes \cite[3.2.4]{Holm} in the context of multiple scales expansions: in a valid expansion its terms must be kept "well-ordered" by magnitude.

Neither the correct time scales $t/\e$, $\e t$ nor the role of $\e\o$ as the more adequate small parameter are apparent from the original system \eqref{CartSlow}, which suggests that it is not the optimal form for studying the behavior of the skidding cart sleigh. In hindsight, the stretched time system \eqref{CartFast} is a more attractive option since it displays the role of $\e\o$ explicitly. We can do even better by multiplying the second equation by $-\e\o$ and setting $\g:=-\e\o\a$. Then the system becomes
\begin{equation}\label{CartEps}
\begin{cases}\b'=\g\\
\g'=-\g-\e^2\o^2\b\,.
\end{cases}
\end{equation}
In this form the system is no longer singularly perturbed in the usual sense (it still is in a relevant technical sense 
\cite[Ch.10]{Ver}), it is apparent that the "correct" small parameter is $\ee:=\e^2\o^2$ rather than $\e$ or $\e\o$, and the regular and compressed time scales $\tau:=t/\e$ and $\ee\tau:=\e\o^2t$ are the "correct" ones from \eqref{CartAnExp}. In particular, it is more effective in investigating the fast motion. Already in the zero order we have 
\begin{equation}\label{CartEpsTau}
\begin{cases}\b=\b_0+\g_0(1-e^{-\tau})\\
\g=\g_0e^{-\tau}\,
\end{cases}
\text{\!\!\!\!\!,\ which translates into\ \ \ } 
\begin{cases}\b=\b_0+\e\o\a_0(1-e^{-t/\e})\\
\a=\a_0e^{-t/\e}\,.
\end{cases}
\end{equation}
This is better than the zero order approximation in \eqref{CartTwoSol}, and one may proceed with the matching aymptotics as we did above. However, in general the correct time scales can not be discerned from the system by such simple manipulations and more complicated techniques are required, e.g. treating different time scales as independent variables and solving partial differential equations involving them, see \cite[3.2.2]{Holm}, \cite[11.4]{Ver}.

\section{Skidding double spear}\label{s4} 

\begin{figure}[H]
\begin{centering}
\includegraphics[scale=1.2]{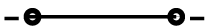}
\par\end{centering}
\caption{\label{DoubleSpear} Skidding double spear. Split off short segments indicate approximate realization of knife-edge constraints by viscous friction.}
\end{figure}

In the cart sleigh both knife-edges were perpendicular to the connecting rod; now we will consider a "double spear", where they are both parallel. The constrained motion is restricted to a straight line and not very interesting, so we start directly with the skidding version, see Fig. \ref{DoubleSpear}.
\noindent The Rayleigh function now is:
$$
F=\frac{c}2\left(\frac{|\dot{r_1}\cdot J(r_1-r_2)|^2}{|r_1-r_2|^2}
+\frac{|\dot{r_2}\cdot J(r_1-r_2)|^2}{|r_1-r_2|^2}\right), 
$$
where $J=\begin{pmatrix}0&-1\\1&0\end{pmatrix}$ is the $90^\circ$ counterclockwise rotation matrix. The vector equations of motion are obtained as before (see \eqref{Deriv} for taking the derivatives):
$$
\begin{cases}m\ddot{r}_1+\l(r_1-r_2)+c\,\ds{\frac{\dot{r_1}\cdot J(r_1-r_2)}{|r_1-r_2|^2}}J(r_1-r_2)=0\\ 
m\ddot{r}_2+\l(r_1-r_2)+c\,\ds{\frac{\dot{r_2}\cdot J(r_1-r_2)}{|r_1-r_2|^2}}J(r_1-r_2)=0\,,\end{cases}
$$
where $\l$ is the Lagrange multiplier for the rigid rod constraint $(\dot{r_1}-\dot{r_2})\cdot(r_1-r_2)=0$. In terms of the center of mass $r:=\frac12(r_1+r_2)$, and the radius vector $r_{12}:=\frac12(r_1-r_2)$ the system is
$$
\begin{cases}
m\ddot{r}_{12}+2\l r_{12}+c\,\ds{\frac{\dot{r}\cdot Jr_{12}}{|r_{12}|^2}\,Jr_{12}}=0\\
m\ddot{r}+c\,\ds{\frac{\dot{r}\cdot Jr_{12}}{|r_{12}|^2}\,Jr_{12}}=0,\text{\ \ }\dot{r}_{12}\cdot r_{12}=0\,.
\end{cases}
$$
As before we set $r_{12} =\rho e^{i\t}$, which reduces the first equation to $m(-\dot{\t}^2+i\ddot{\t})+2\l+ic\dot{\t}=0$, so with $\o:=\dot{\t}$ we have $\l=m\o^2/2$ and $m\dot{\o}+c\o=0$. In contrast to the cart sleigh, the rotational angular velocity is not constant in this case. 

Setting $\dot{r}=zr_{12}=(\a+i\b)r_{12}$ reduces the second equation to 
$$
m(\dot{\a}-\b\o)+im(\dot{\b}+\a\o)+ic\b=0\,,
$$
and with $\e:=m/c$ we obtain
\begin{equation}\label{SpearEq}
\begin{cases}\dot{\a}=\o\b\\
\e\dot{\b}=-\b-\e\o\a\\
\e\dot{\o}=-\o\,.
\end{cases}
\end{equation}
The last equation decouples from the other two, and the solution to it is $\o=\o_0e^{-t/\e}$. We therefore have 
\begin{equation}\label{Solr12}
r_{12}(t)=\rho e^{i(\t_0+\int_0^t\o(t)\,dt)}
=\rho e^{i\left(\t_0+\e\o_0(1-e^{-t/\e})\right)},
\end{equation}
which means that $r_{12}(t)$ quickly settles into the limit value of $\rho e^{i(\t_0+\e\o_0)}$. Moreover, \eqref{SpearEq} reduces to a non-autonomous system in two variables:
\begin{equation}\label{SpearTimeEq}
\begin{cases}
\dot{\a}=\o_0e^{-t/\e}\b\\
\e\dot{\b}=-\b-\e\o_0e^{-t/\e}\a\,.
\end{cases}
\end{equation}
The fast time scale $t/\e$ is now explicitly present in the coefficients, which makes the substitution $\tau:=t/\e$ all the more natural, so that with primes denoting $\tau$ derivatives
\begin{equation}\label{SpearTimeFast}
\begin{cases}
\a'=\e\o_0e^{-\tau}\b\\
\b'=-\b-\e\o_0e^{-\tau}\a\,.
\end{cases}
\end{equation}
With the benefit of past experience we set right away $\g:=\e\o_0\b$ and $\ee:=\e^2\o_0^2$ to finally obtain 
\begin{equation}\label{SpearTimeEeps}
\begin{cases}
\a'=e^{-\tau}\g\\
\g'=-\g-\ee\,e^{-\tau}\a\,.
\end{cases}
\end{equation}
The regular expansions to the first order are $\a=\a^0+\ee\a^1+\dots$, $\g=\g^0+\ee\g^1+\dots$, and upon the substitution 
into \eqref{SpearTimeEeps} we have
\begin{equation}\label{SpearFastExp}
\begin{cases}
\a^{0\,\prime}+\ee\a^{1\,\prime}+\dots=e^{-\tau}(\g^0+\ee\g^1+\dots)\\
\g^{0\,\prime}+\ee\g^{1\,\prime}+\dots=-\g^0-\ee\g^1-\dots-\ee\,e^{-\tau}(\a^0+\ee\a^1+\dots)\,.
\end{cases}
\end{equation}
In the zero order in $\ee$ this gives 
\begin{equation*}
\begin{cases}
\a^{0\,\prime}=e^{-\tau}\g^0\\
\g^{0\,\prime}=-\g^0
\end{cases}
\end{equation*} with solutions
\begin{equation}\label{SpearFastExp0}
\g^0=\g_0\,e^{-\tau},\ \ \a^0=\a_0 +\frac{\g^0}2(1-e^{-2\tau}).
\end{equation}
Here $\g_0:=\g(0)$ and $\a_0:=\a(0)$ are the initial values, which we substituted directly because $\tau$ is already the time scale of the fast motion. In the first order in $\ee$ we have 
\begin{equation}\label{SpearFastEq1}
\begin{cases}
\a^{1\,\prime}=e^{-\tau}\g^1\\
\g^{1\,\prime}=-\g^1-e^{-\tau}\a^0\,,
\end{cases}
\end{equation}
and the initial values are $\a^1(0)=\g^1(0)=0$ since $\a_0$, $\g_0$ are already accounted for in the zero order. 
Therefore, the first order solutions are:
\begin{equation}\label{SpearFastExp1}
\begin{cases}
\a^{1}=-\frac{\a_0}4(1-e^{-2\tau})+\frac12(\a_0+\frac{\g^0}2)\tau e^{-2\tau}-\frac{\g^0}{16}(1-e^{-4\tau})\\
\g^{1}=-(\a_0+\frac{\g^0}2)\tau e^{-\tau}+\frac{\g^0}4e^{-\tau}(1-e^{-2\tau})\,,
\end{cases}
\end{equation}
Note the presence of secular terms, which are however suppressed by negative exponents this time. Taking $\tau\to\infty$ we obtain an approximation of the values "after" the fast motion: $\g=0$ and
\begin{equation*}
\a=\a_0+\frac{\e\o_0}2\,\b_0-\frac{\e^2\o_0^2}4\,\a_0-\frac{\e^3\o_0^3}{16}\,\b_0+O(\e^3)\,.\\
\end{equation*}
The compressed time scale is $t_1=\ee\tau$ with $\frac{d}{d\tau}=\ee\frac{d}{dt_1}$. Still using primes for derivatives we have
\begin{equation*}
\begin{cases}
\ee\,\a^{\prime}=e^{-t_1/\ee}\g\\
\ee\,\g^{\prime}=-\g-\ee\,e^{-t_1/\ee}\a
\end{cases}
\end{equation*}
or substituting expansions:
\begin{equation}\label{SpearSlowExp}
\begin{cases}
\ee(\a^{0\,\prime}+\ee\a^{1\,\prime}+\dots)=e^{-t_1/\ee}(\g^0+\ee\g^1+\dots)\\
\ee(\g^{0\,\prime}+\ee\g^{1\,\prime}+\dots)=-\g^0-\ee\g^1-\dots-\ee\,e^{-t_1/\ee}(\a^{0}+\ee\a^{1}+\dots)\,,
\end{cases}
\end{equation}
In the zero order we get $\g^0=0$, which matches the fast motion. In the first order
$$
\begin{cases}
\ee\,\a^{0\,\prime}=e^{-t_1/\ee}\g^1\\
\ee\,\g^{0\,\prime}=-\g^1-e^{-t_1/\ee}\a^{0}\,,
\end{cases}
$$  
hence $\g^1=-e^{-t_1/\ee}\a^{0}$ and $\a^{0\,\prime}=-e^{-2t_1/\ee}\a^{0}$, so $\a^{0}=ae^{\frac{\ee}2e^{-2t_1/\ee}}$ for some constant $a$. When $t_1=0$ we have $\a^{0}=ae^{\frac{\ee}2}=\a_0+\frac{\g_0}2$, and matching with the fast motion gives 
$\a=(\a_0+\frac{\g_0}2)e^{-\frac{\ee}2(1-e^{-2t_1/\ee})}=(\a_0+\frac{\g_0}2)e^{-\frac{\ee}2(1-e^{-2\tau})}$. As seen from the last expression no new time scale appears in the answer even when we try to "force" it. This is in contrast to the case of the cart sleigh, and may suggest that the regular expansion in $\tau$ already works for all times uniformly. But even in this example it does not, albeit for a different reason than for the cart sleigh. To see why let us again compare to the analytic solution.

We start by transforming \eqref{SpearTimeEeps} into a single second order equation for $\a$:
\begin{equation}\label{SpearSingleTau}
\a''+2\a'+\ee\,e^{-2\tau}\a=0\,.
\end{equation}
The substitution $x:=\ee^{1/2}\,e^{-\tau}$ transforms it further into $x\a_{xx}-\a_x+x\a=0$, which is a well-known equation in mathematical physics. It turns into one known even better after the substitution $\a:=xu$, namely into 
\begin{equation}\label{SpearSingleBes}
x^2u_{xx}+x\,u_x+(x^2-\nu^2)u=0
\end{equation}
with $\nu=1$. Equation \eqref{SpearSingleBes} is the equation of cylindrical (Bessel) functions \cite{Kor}. Their appearance is somewhat surprising here because they relate to behavior in time, and have nothing to do with geometry. The two linearly independent solutions to \eqref{SpearSingleBes} are the Bessel functions of the first and the second kind $J_\nu(x)$ and $Y_\nu(x)$, so the general solution can be written as $\a(x)=ax\,J_1(x)+bx\,Y_1(x)$ with $x=\ee^{1/2}\,e^{-\tau}$, and arbitrary constants $a, b$.

To solve for $a, b$ in terms of initial values we will use some standard properties of Bessel functions. Namely, the derivative formulas $\big(x^\nu\,J_\nu(x)\big)'=x^\nu\,J_{\nu-1}(x)$, $\big(x^\nu\,Y_\nu(x)\big)'=x^\nu\,Y_{\nu-1}(x)$, and the Wronskian identity:
\begin{equation}\label{BesWron}
\begin{vmatrix}J_\nu(x)&Y_\nu(x)\\J'_\nu(x)&Y'_\nu(x)\end{vmatrix}
=\begin{vmatrix}J_\nu(x)&Y_\nu(x)\\J_{\nu-1}(x)&Y_{\nu-1}(x)\end{vmatrix}=\frac2{\pi x}\,.
\end{equation}
Note that $\frac{d}{d\tau}\a|_{\tau=0}=e^{-\tau}\g|_{\tau=0}=\g(0)=\g_0$, and $\frac{d}{d\tau}=-x\frac{d}{dx}$, therefore 
$\frac{d\a}{d\tau}(0)=-x\frac{d\a}{dx}|_{x=\ee^{1/2}}=-\ee^{1/2}\a_x(\ee^{1/2})$.
We also have 
$$
\a_x=a\big(x J_1(x)\big)'+b\big(x Y_1(x)\big)'=ax\,J_0(x)+bx\,Y_0(x),
$$
so the system for $a,b$ is
$$
\begin{cases}
\ee^{1/2}J_1(\ee^{1/2})a+\ee^{1/2}Y_1(\ee^{1/2})b=\a_0\\
\ee^{1/2}J_0(\ee^{1/2})a+\ee^{1/2}Y_0(\ee^{1/2})b=-\g_0\,\ee^{1/2}\,.
\end{cases}
$$  
Using Cramer's rule and \eqref{BesWron}: 
\begin{equation*}
\begin{pmatrix}a\\b\end{pmatrix}
=\frac{\pi}2\begin{pmatrix}Y_0(\ee^{1/2})&-Y_1(\ee^{1/2})\\-J_0(\ee^{1/2})&J_1(\ee^{1/2})\end{pmatrix}
\begin{pmatrix}\a_0\\-\ee^{1/2}\g_0\end{pmatrix}\,.
\end{equation*}
Since $\ee=\e^2\o^2_0$ and $\g=\e\o_0\b$ we have $\ee^{-1/2}\g_0=\sign(\o_0)\b_0$, and $\sign(\o_0)\ee^{1/2}\g_0=\e\o_0$. 
Thus,
\begin{multline*}
\a=\frac{\pi}2\Big(\a_0\,Y_0(\ee^{1/2})+\sign(\o_0)\b_0\,Y_1(\ee^{1/2})\Big)\,
\ee^{1/2}\,e^{-\tau}J_1(\ee^{1/2}\,e^{-\tau})\\
-\frac{\pi}2\Big(\a_0\,J_0(\ee^{1/2})+\sign(\o_0)\b_0\,J_1(\ee^{1/2})\Big)\,
\ee^{1/2}\,e^{-\tau}Y_1(\ee^{1/2}\,e^{-\tau})
\end{multline*}
\vspace{-2.5em}
\begin{multline}\label{SpearAnSol}
\b=-\frac{\pi}2\Big(\a_0\,Y_0(\ee^{1/2})+\sign(\o_0)\b_0\,Y_1(\ee^{1/2})\Big)\,
\ee^{1/2}\,e^{-\tau}J_0(\ee^{1/2}\,e^{-\tau})\\
+\frac{\pi}2\Big(\a_0\,J_0(\ee^{1/2})+\sign(\o_0)\b_0\,J_1(\ee^{1/2})\Big)\,
\ee^{1/2}\,e^{-\tau}Y_0(\ee^{1/2}\,e^{-\tau})
\end{multline}
\begin{figure}[H]
\begin{centering}
\includegraphics[scale=0.5]{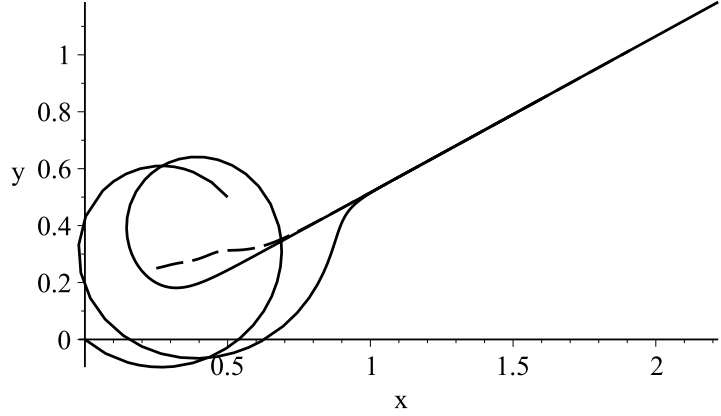}
\par\end{centering}
\caption{\label{p2spearhighomega} Skidding double spear moving. Solid lines are the trajectories of the knife-edges, dashed line is the trajectory of the center of mass.}
\end{figure}
Figure \ref{p2spearhighomega} shows the trajectories of the masses and of the center of mass for motion with the  relatively large initial velocity to highlight the transient phase. As expected from \eqref{SpearFastExp0} on the time scale of $t\sim\e$ after some spinning the spear settles into the nearly rectilinear motion characteristic of the constrained system. In other words, the frictional trajectory shadows a constrained one in this case.

Since formulas \eqref{SpearAnSol} are rather obscure we will make use of the following Taylor expansions at $x=0$:
\begin{align}\label{BesTaylor}
J_0(x)&=\sum_{k=0}^\infty\frac{(-1)^k}{(k!)^2}\left(\frac{x}2\right)^{2k};
\ \ \ \ \ \ \ \ \ \,J_1(x)=\frac{x}2\sum_{k=0}^\infty\frac{(-1)^k}{k!(k+1)!}\left(\frac{x}2\right)^{2k};\notag\\
Y_0(x)&=\frac2\pi\left[J_0(x)\Big(\ln\frac{|x|}2+C\Big)-\sum_{k=1}^\infty\frac{(-1)^k}{(k!)^2}
\Big(1+\dots+\frac1k\Big)\left(\frac{x}2\right)^{2k}\right];\\
Y_1(x)&=\frac2\pi\left[-\frac1x+J_1(x)\Big(\ln\frac{|x|}2+C\Big)\right.\notag\\
&\hspace{6em}\left.-\frac{x}4\left(1+\sum_{k=1}^\infty\frac{(-1)^k}{k!(k+1)!}
\Big(2+\dots+\frac2k+\frac1{k+1}\Big)\left(\frac{x}2\right)^{2k}\right)\right].\notag
\end{align}
In \eqref{BesTaylor} $C$ is the Euler constant. Clearly $J_0,Y_0$ are even and $J_1,Y_1$ are odd, so \eqref{SpearAnSol} will not change if all entries of $\ee^{1/2}$ are replace by $\e\o_0$, and $\sign(\o_0)$ is removed. Expanding into the powers of 
$\ee^{1/2}$ with this in mind we obtain
\begin{align}\label{SpearAnExp}
\a&=\a_0+\frac{\b_0}2(1-e^{-2\tau})\e\o_0-\frac{\a_0}2\big({\textstyle\frac12}(1-e^{-2\tau})-\tau e^{-2\tau}\big)\e^2\o_0^2+O(\e^3\o_0^3)\notag\\
\b&=\b_0e^{-\tau}-\a_0\tau e^{-\tau}\e\o_0+\frac{\b_0}4\big((1-2\tau)e^{-\tau}+e^{-3\tau}\big)\e^2\o_0^2
+O\big(\e^3\o_0^3\ln|\e\o_0|\big)\,.
\end{align}
Note that the logarithmic terms appearing in \eqref{BesTaylor} cancel out in \eqref{SpearAnExp} to the order displayed. Taking into account that $\g=\e\o_0\b$, $\g_0=\e\o_0\b_0$ we also see that it matches the regular expansion obtained in 
\eqref{SpearFastExp0} and \eqref{SpearFastExp1} to this order. 
\vspace{1em}
\begin{figure}[H]
\vspace{-0.3in}
\begin{centering}
(a) \includegraphics[scale=0.3]{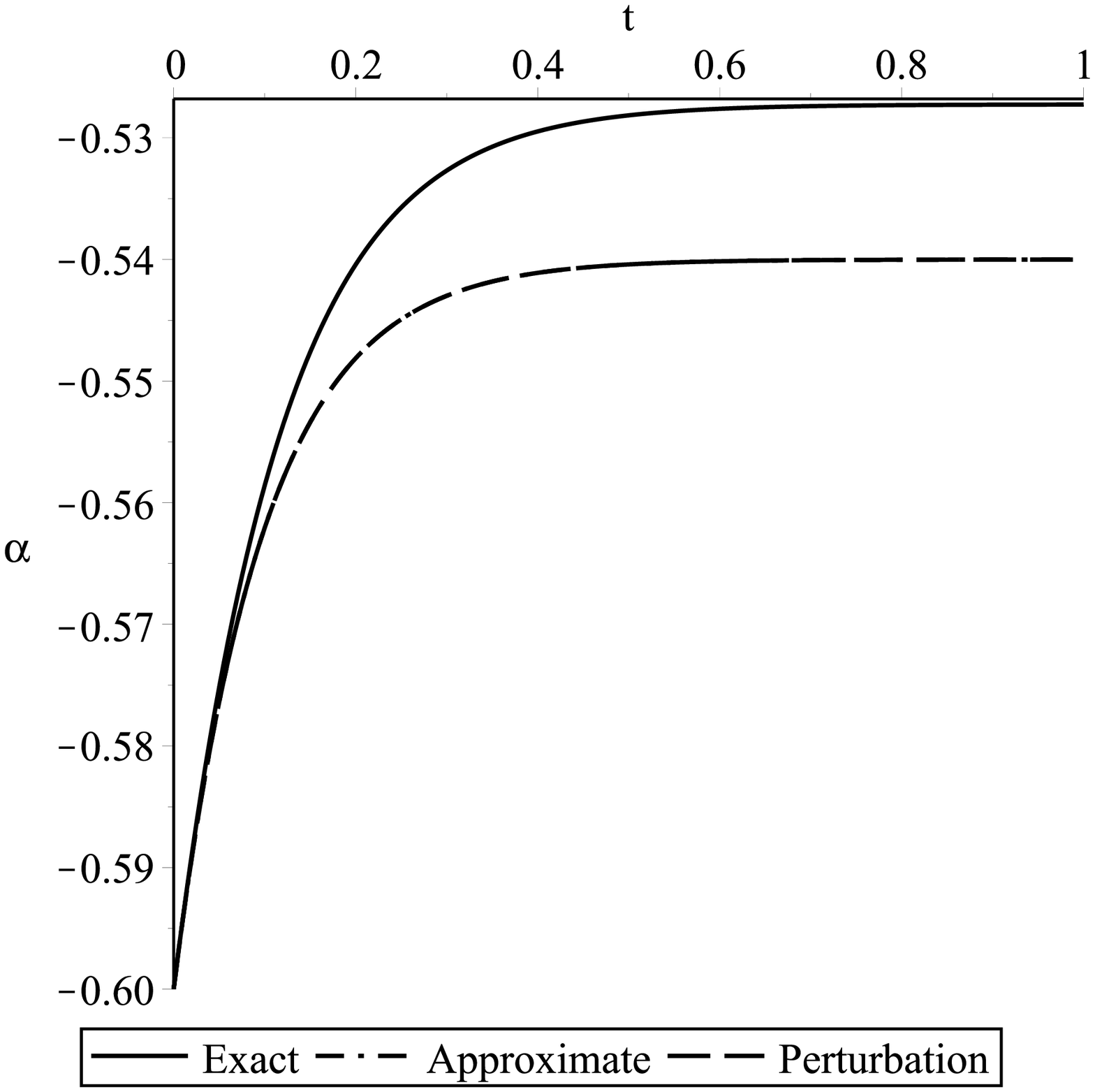}
(b) \includegraphics[scale=0.3]{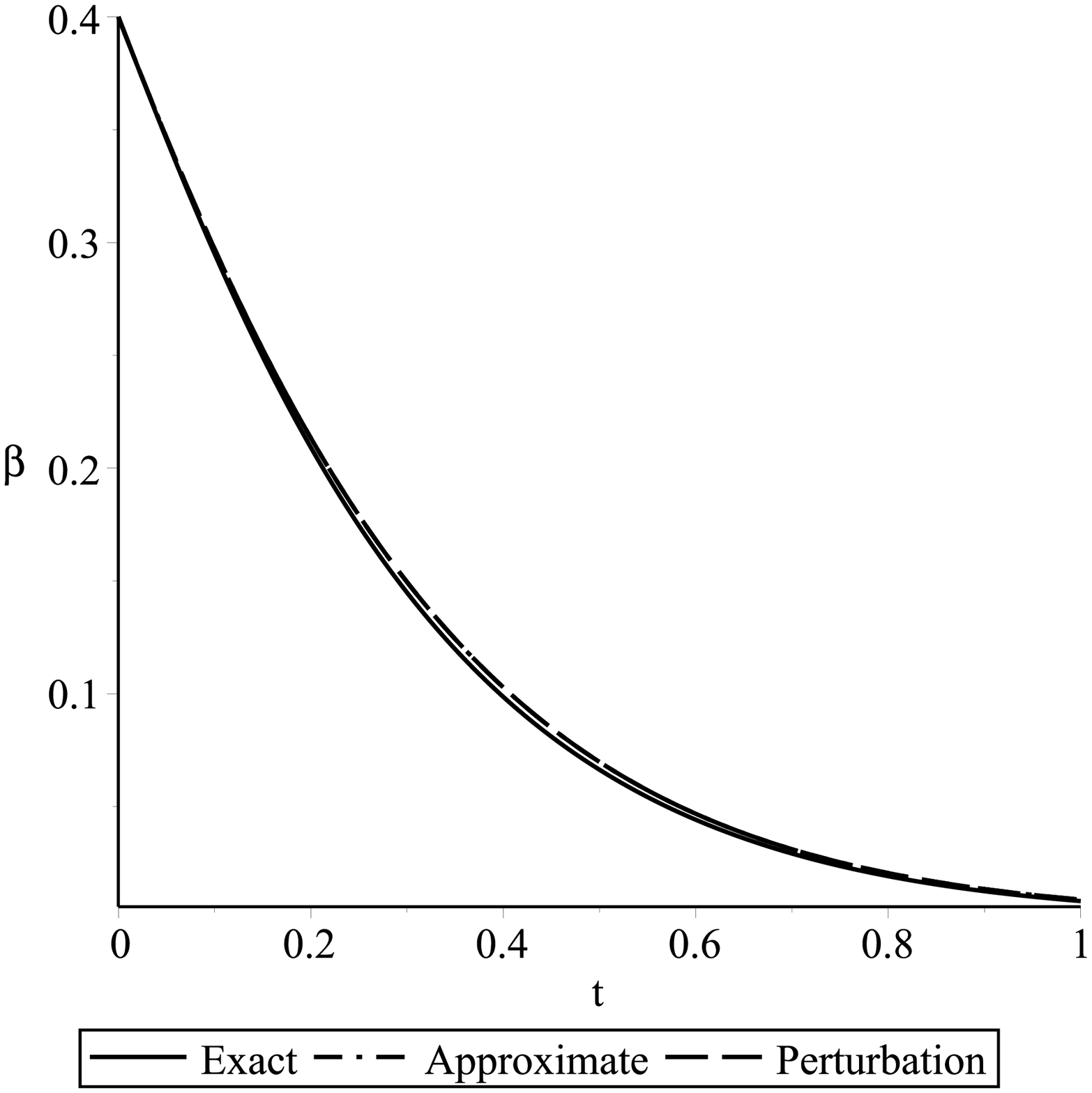} 
\par\end{centering}
\caption{\label{2alpha} Exact, perturbative and approximate solution for $\a$ (a) and $\b$ (b) for initial conditions 
$r_1(0)=0$, $r_2(0)=2+2i$, $\dot{r}(0)=1+i/5$, $\o=3/2$ and $\e=1/5$.}
\end{figure}
Due to the exponential suppressing factor in $x=\ee^{1/2}e^{-\tau}$ the uniform convergence of \eqref{SpearAnExp} in time is implied by the mere local convergence of the Taylor series in $x$ at $0$. Keep in mind however that logarithmic terms do appear in higher orders, while they are by definition absent from the regular expansion, which therefore does not converge uniformly as mentioned earlier, despite the exponential suppression of secular terms. The challenge however is to extract such information from the equations of motion \eqref{CartSlow} and \eqref{SpearEq}, rather than from the analytic solutions. On Fig.\ref{2alpha} graphs of the regular perturbative, and the analytic expansions (to the second order) coincide, but both slightly but visibly deviate from the analytic solution for large times.

\section{Conclusions}\label{s5} 

We showed that plane dumbbells provide useful toy models for analyzing viscous realizations of non-holonomic systems, both for their technical simplicity, and richness of displayed phenomena. Analysis of their motion was shown to be greatly aided by treating positions of the masses as complex numbers, which leads to an elementary version of symmetry reduction to suitable generalized coordinates. The reduced equations of motion are more tractable, and sometimes linear and/or can be solved analytically. 

Even when an analytic solution is unavailable methods of singular perturbation theory can be used to analyze large viscosity approximations of non-holonomic motion quantitatively. This is because the ratio $m/c$ serves as a natural small parameter. A number of approximating methods have been developed for analyzing systems with a small parameter. Let us briefly survey some of them to provide context. 

Perhaps the most straightforward method is the regular (Taylor) expansion of the solution in the powers of the parameter with iterative determinations of coefficients from the equations and initial conditions. Unfortunately, the resulting expansion works poorly in the long term, and not at all for singularly perturbed systems. A modification of it, known as the "two timing" expansion, was theoretically justified by O'Malley and Vasil'eva in 1960s based on Tikhonov's theory of singularly perturbed systems \cite[8.3]{Ver}. We used Vasil'eva's procedure for determining the coefficients of these expansions, known as "matching asymptotics". Although our examples are, in the end, linear, the generalization to non-linear systems is straightforward. The method is quite attractive for analyzing transient motions because it is straightforward to apply, and provides good accuracy. Nonetheless, the method's utility is limited for a number of reasons: appearance of more than two time scales, especially in non-linear systems, secular terms, logarithmic behavior, etc. Moreover, equations of motion can be misleading, the time scales "apparent" from them may not be the time scales that best represent the system's behavior.

For oscillatory systems some refinements of the expansion methods aimed at long term approximations were developed, such as the Poincare-Lindstedt method and the method of averaging \cite[3.1]{Holm}, \cite[10.1]{Ver}. Unfortunately, they are specific to oscillatory problems, as they are based on "averaging out" oscillations when determining coefficients to eliminate the secular terms. Both methods can be seen as shortcuts for the general method of multiple time scales, where the latter are treated as independent variables to be determined by solving partial differential equations \cite[3.2.2]{Holm}, \cite[11.4]{Ver}. The method is quite cumbersome even in linear, and multiply so in non-linear, cases.

One of the main reasons the matched perturbative expansions do not represent the motion on longer time scales is the appearance of the so-called secular terms, whose exponents are multiplied by polynomials. As we saw, in the examples with dumbbells, viscously perturbed systems are quite peculiar in this regard. Neither the unperturbed (constrained) solutions nor their perturbations are oscillatory, as in the paradigmatic examples of secular terms in celestial mechanics and electric engineering. We believe that the phenomenon is general and calls for a new approach to elimination of secular terms, specific to this context, which can serve as shortcuts to be solving partial differential equations for appropriate time scale.

In a different direction, it turned out that the attractive idea of using viscous realization to define "generalized solutions" to constrained systems for initial values not conforming to the constraints, does not work in general. One problem is that the frictional trajectories do not always "shadow", i.e. approximate uniformly in time, any constrained trajectory, including those obtained by replacing non-conforming initial values by conforming ones "at the end" of the fast motion. The overall picture that emerges is this. The theory implies that on a fixed time interval viscous motions converge to constrained motions when the damping coefficient $c$ goes to infinity \cite{Bren, Kar}. One can split the viscous motion into initial fast motion, during which the system quickly approaches the manifold of constrained motions, and subsequent slow motion in its neighborhood. One might expect that during the latter the viscous motion shadows some non-holonomic trajectory in the constrained manifold.

However, as we saw in the example of the skidding cart sleigh, just because a system stays close to the constrained manifold does not mean that it stays close to any particular trajectory in it, i.e. there may not be any solution to the constrained equations of motion that viscous solutions converge to. Indeed, one can see from \eqref{CartAnSol} that for any $\e>0$ we have $\b_\e(t)\xrightarrow[t\to\infty]{}0$, whereas in the limit $\b=\b_0=\const$. In other words, no matter how small $\e$ is, for $t$ large enough the viscous trajectory will come apart from any constrained trajectory, despite remaining close to the constrained manifold at all times. In fact, the global behavior of the system with viscous friction is qualitatively different from that of the constrained system. In the former the sleigh asymptotically comes to a stop, whereas in the latter it keeps going forever. Therefore, in general there is no constrained solution that frictional solutions approach in the limit of infinite damping starting from non-conforming initial values. This phenomenon is called drift or creep dynamics \cite{Eld}.

This means that the fast/slow motion picture does not always tell the whole story of non-holonomic constraints realized by viscous friction. There is an additional complication in that the manifold of constrained motions is typically non-compact, e.g. for the skidding sleigh it is $\{(\a,\b,\o)\ |\ \a=0\}$ (at the level of reduced variables), and we generally expect a non-compact range of motions for constrained systems. Perturbation theory for non-compact invariant manifolds is developed in \cite{Eld}. However, while such manifolds persist under perturbations, the same can not be said about global attractors which can change abruptly, and it is their structure that determines asymptotic behavior. And this can happen even when the configuration space is compact.
\begin{figure}[H]
\begin{centering}
\includegraphics[scale=1.2]{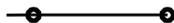}
\par\end{centering}
\caption{\label{ChapSleigh} Chaplygin sleigh dumbbell.}
\end{figure}
We expect the same conclusions to hold for the viscoelastic realization of constraints considered in \cite{Dep}, where convergence is also proved only for $t\sim O(1)$.

In the light of this let us look back at the classical example of the Chaplygin sleigh, which is equivalent to the dumbbell in Fig. \ref{ChapSleigh}. Carath\'eodory \cite{Car} reasoned heuristically that for arbitrarily large damping the trajectories of the skidding sleigh differ dramatically from the ideal ones. Fufaev \cite{Fuf}, see also \cite[IV.3]{NF}, gave a fast/slow motion analysis of the situation explaining that the viscous motion does in fact approximate the constrained one shortly after $t=0$ (although not at $t=0$), which undermined Carath\'eodory's reasoning. Fufaev did not, however, analyze long term behavior of the skidding sleigh other than to say that it stays close to the constrained manifold. As we saw for simpler dumbbells, the two motions may yet come apart. Could Carath\'eodory have been right, if only for a wrong reason? 

Numerical simulations in \cite{Eld} suggest that for velocities the answer is negative, and there is shadowing just as in the case of the double spear (Figure 8 in \cite{Eld} shows that the asymptotic directions of motion differ, and so there is no shadowing for position variables). But the Carath\'eodory-Fufaev sleigh equations are more involved, and analytic solutions to them are not available. It would be interesting to find the answer definitively, and even more desirable to find a general approach to answering such questions without the recourse to analytic solutions. 
\\

\textbf{Acknowledgement}: The authors are grateful to Jaap Eldering for multiple suggestions and corrections during the preparation of the paper.

\end{document}